\documentclass[aps,prb,amsmath,amssymb,superscriptaddress,twocolumn]{revtex4-1}
\usepackage{CJKutf8}
\usepackage{graphicx}
\usepackage{dcolumn}
\usepackage{amsmath}
\usepackage{bm}
\usepackage{physics}
\usepackage{verbatim}
\usepackage{color}
\usepackage{tabularx}
\usepackage{xcolor}
\usepackage{cancel}
\usepackage{hyperref}
\usepackage{siunitx}
\usepackage{gensymb}

\usepackage{xr}
\definecolor{gold}{rgb}{0.83, 0.69, 0.22}

\newcommand\COMMENTED[1] {}

\newcommand{\bs}{\boldsymbol}

\begin{document}
\begin{CJK*}{UTF8}{gbsn}
\title{Engineering Hubbard models with gated two-dimensional moiré systems}

\author{Yiqi Yang (杨怡奇)}
\email{yyang25@wm.edu}
\affiliation{Department  of  Physics,  College  of  William  \&  Mary,  Williamsburg,  Virginia  23187, USA}
\author{Yubo Yang (杨煜波)}
\affiliation{Department of Physics and Astronomy, Hofstra University, Hempstead, New York 11549, USA}
\author{Miguel A. Morales}
\affiliation{Center for Computational Quantum Physics, Flatiron Institute, 162 5th Avenue, New York, New York, USA}
\author{Shiwei Zhang}
\affiliation{Center for Computational Quantum Physics, Flatiron Institute, 162 5th Avenue, New York, New York, USA}

\begin{abstract}
Lattice models are powerful tools for studying strongly correlated quantum many-body systems, but their general lack of exact solutions motivates efforts to simulate them in tunable platforms.
Recently, a promising new candidate has emerged for such platforms from
two-dimensional materials. A subset of moiré systems can be effectively described as a two-dimensional electron gas (2D EG) subject to a moiré potential, with electron-electron interactions screened by nearby metallic gates. In this paper, we investigate the realization of lattice models in such systems.
We show that, by controlling the gate separation, a 2D EG in a square moiré potential 
can be systematically tuned into a system whose
ground state exhibits orders analogous to those of the square lattice Hubbard model, including the stripe phase. Furthermore, we study how variations in gate separation and moiré potential depth 
affect the ground-state orders. 
A number of antiferromagnetic phases, as well as a ferromagnetic phase and a paramagnetic phase, are identified. 
We then apply our quantitative downfolding approach to triangular moiré systems closer 
to current experimental conditions, compare them with the square lattice parameters 
studied, and outline 
routes for experimental realization of the phases. 
\end{abstract}
\maketitle
\end{CJK*}

\section{Introduction}
Lattice models provide a fundamental framework for understanding the physics of strongly correlated quantum many-body systems. However, exact solutions to these models are rarely available, motivating experimental efforts to construct quantum simulators that can directly emulate lattice Hamiltonians. 
A prime example has been with neutral atoms in optical lattices~\cite{Xu_Nature2025}.
Traditionally, model engineering has been 
challenging in condensed matter. However, 
recent advances in two-dimensional materials have presented exciting new opportunities.
A promising platform for such simulations involves stacking and twisting multiple layers of two-dimensional (2D) materials. When confined in two dimensions, electron kinetic energy is suppressed and the electron-electron interaction is enhanced owing to the reduced screening effects from other electrons. In the presence of a moiré potential, the electron kinetic energy is further suppressed. The relative interaction strength can be flexibly tuned by adjusting the twist angle~\cite{Wu_PRL2018} and the surrounding dielectric environment~\cite{Rosner_PRB2015,Loon_NPJ2023}.
Often metallic gates are present in these systems to tune carrier density, and the gate separation offers additional control over both the range and magnitude of interactions~\cite{Yang_PRB2025,Valenti_PRL2025}. 

Realizing lattice models with the 2D materials platform has received much recent attention.  
Various lattice models, such as the triangular~\cite{Wu_PRL2018}, honeycomb, and kagome lattices~\cite{Angeli_PNAS2021}, have been theoretically proposed to be realizable using heterobilayers or homobilayers of transition metal dichalcogenides (TMDs). More recently, the simulation of the square lattice models has been proposed in twisted bilayer $\text{FeSe}$~\cite{Eugenio_SciPostPhys2023}, $\text{GeX}/\mathrm{SnX} \mathrm{(X}=\mathrm{S,Se)}$~\cite{Xu_Arxiv2024}, carbon allotrope $\text{C}_{568}$~\cite{Kariyado_PRB2025}, and more general homobilayers~\cite{Eugenio_PRL2025}. 
Experimental realization of the triangular lattice model has been achieved by stacking a monolayer of $\mathrm{WSe}_2$ on top of a monolayer of $\mathrm{WS}_2$~\cite{Tang_Nature2020,Regan_Nature2020,Tang_NN2023}.

Quantitative guidance for the realization of these models has been rare, however. 
In large part this shortage is not surprising. The systems involved tend to be strongly correlated, with long-range Coulomb interactions which are screened by 
metallic gates. They place high demands 
on accurate theoretical and computational treatment.
Deriving effective model Hamiltonians for such systems requires proper downfolding.  Since the engineered system will necessarily have 
differences from the targeted, idealize model, it is important to have a 
gauge on whether this 
difference can alter the desired property in any fundamental way. 
Reliable quantitative predictions of the properties of the cleanest models are already 
challenging, and treating the engineered system to discern the difference can be
even more difficult.

\begin{figure*}
\includegraphics[width=1.0\textwidth]{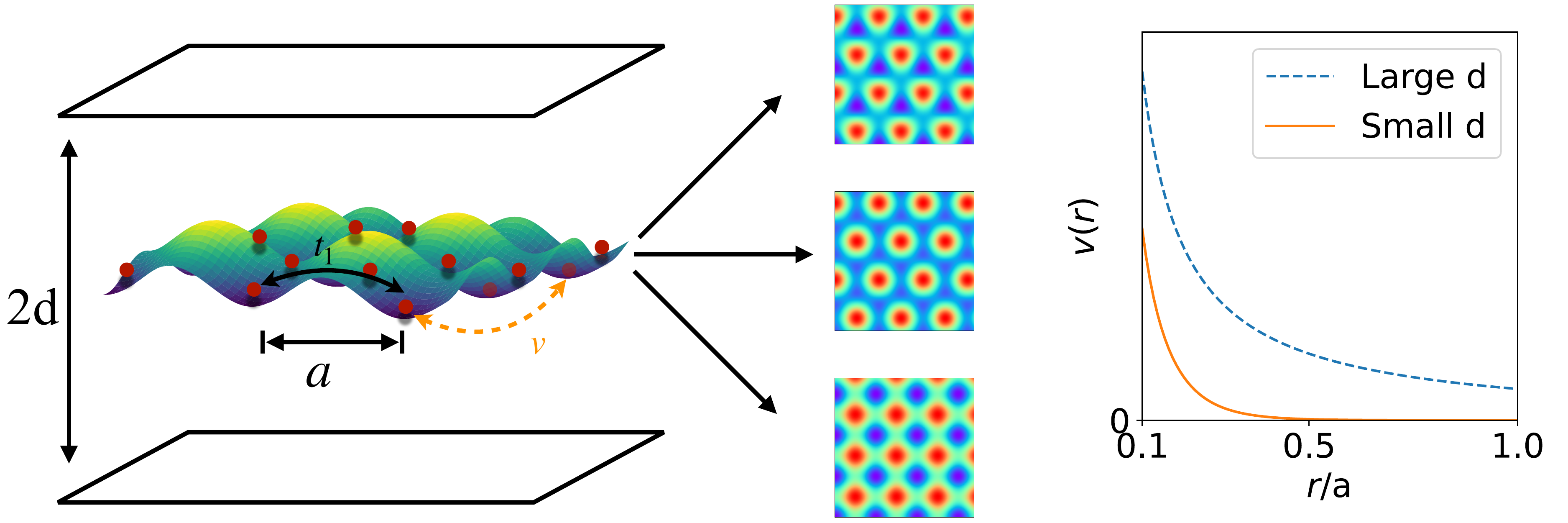}
\centering
\caption{The schematic of a dual-gate screened 2D EG in a moiré potential. In the left panel, electrons denoted by the red dots are confined in a 2D plane with a moiré potential. The squares above and below the 2D EG represent metallic gates, which screen the electron-electron interaction. The middle three panels are moiré potential of various geometries, where the blue color represent potential valleys and the red color represent potential peaks. From top to bottom, the potential minima form triangular, honeycomb, and square lattices, respectively. The right panel illustrates that the dual-gate screening shortens the range of the electron-electron interaction.}
\label{fig:schematic}
\end{figure*}

In this paper, we focus on the engineering of Hubbard models with a 2D EG that lives in a moiré potential.
The Hubbard model, originally introduced to study magnetism in materials with narrow $3d$ bands~\cite{Hubbard_PRSLA1963}, has been extensively explored due to its simple Hamiltonian, rich phase diagram, and relevance to high-$T_c$ superconductivity~\cite{Qin_ARCMP2022}. Realizing the square Hubbard model in moiré materials would provide a valuable experimental platform to complement existing theoretical studies, for example by enabling transport measurements to probe superconductivity. 
In Fig.~\ref{fig:schematic} we provide a schematic overview of the system we study and the basic idea of our work. 

\begin{figure}[h]
\includegraphics[width=0.5\textwidth]{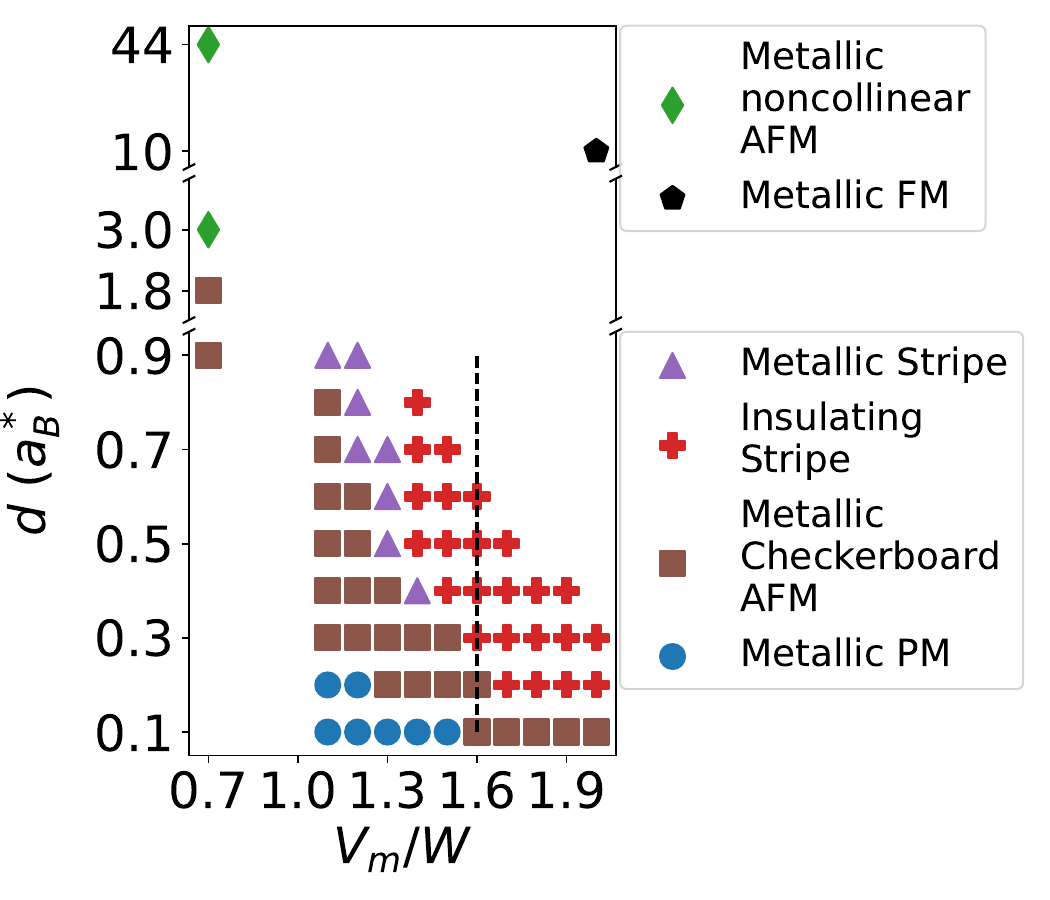}
\centering
\caption{Phase diagram for the square moiré potential at $1/8$ hole doping} and $r_s=5\; a_B^*$, as a function of the gate distance $d$ and potential depth $V_m$, normalized by the kinetic energy scale $W=1/r_s^2$. In the legend, AFM, FM, and PM stand for  antiferromagnetic, ferromagnetic, and paramagnetic, respectively.
The vertical dashed black line indicates parameter sets we downfold to lattice models for comparison in Sec.~\ref{sec:results}.
\label{fig:phaseDiagram}
\end{figure}

We study the ground state order of the 2D EG system as a function of the moiré potential depth and the gate separation. We employ a density-functional specifically parametrized for this system~\cite{Yang_PRB2025}, and perform density-functional theory (DFT) calculations under the 
local-density approximation (LDA), 
more specifically local spin-density approximation (LSDA).
We scan the interaction range, strength, and density/filling systematically to
identify ground-state orders.
With a square moiré potential and gate separation much smaller than the moiré lattice constant, we find ground-state orders that closely resemble those found in the square lattice Hubbard model, including the stripe phase~\cite{Xu_PRR2022,Xu_JPCM2011}. 
Away from this Hubbard regime, the system exhibits alternative ground-state orders. A phase diagram for $1/8$ hole doping from half-filling is shown in Fig~\ref{fig:phaseDiagram}, and discussed in further detail in Sec.~\ref{ssec:Hub-doped}.

To gain intuitive insight into the underlying lattice model and facilitate direct comparison with experiments, 
we also perform systematic downfolding using Wannier functions to map the 2D EG in both triangular and square moiré potentials onto effective lattice models. We find that in the parameter regime where the stripe phase emerges, both longer-range hopping and inter-orbital interactions are negligible. The ratio of intra-orbital interaction $U_0$ to nearest-neighbor hopping $t_1$ in this regime are consistent 
with those associated with the stripe phase in the square Hubbard model. Notably, the critical $U_0/t_1$ required to stabilize the stripe phase in the square moiré system is significantly smaller than that in triangular moiré systems realized in current experiments.

The remainder of this paper is organized as follows. In Sec.\ref{sec:model}, we describe our system 
in detail. Section\ref{sec:method} outlines the computational methods employed in this work. In Sec.\ref{sec:results}, we present our main results. We conclude and share some outlooks in Sec.~\ref{sec:Conclusions}.

\section{Dual-gate screened 2D EG in moir\'e potential}\label{sec:model}
We consider a 2D EG that lives in a moiré potential, as shown schematically in Fig~\ref{fig:schematic}.
Besides intrinsic interest, such a system can model the low energy physics of heterobilayer TMDs, where the valley degrees of freedom is mapped to the electron spins in a 2D EG, in the presence of spin-valley locking ~\cite{Wu_PRL2018,Yang_PRL2024,YangYubo_arxiv2024}.  
We use the effective Bohr radius $a_B^*=a_B\times(\epsilon/\epsilon_0)/(m^*/m_e)$ and effective Hartree $E_h^*=E_h\times(m^*/m_e)/(\epsilon/\epsilon_0)^2$ as length and energy units~\cite{Kohn_Donor_1955}, respectively, where $m^*$ is the effective mass of the charge carriers and $\epsilon$ the relative permittivity (dielectric constant) of the dielectric spacers. The Hamiltonian of the system is
\begin{align}
\label{eq:H-2deg-moire}
\hat{H} =& -\dfrac{1}{2}\sum\limits_{i=1}^N \nabla_i^2 +\dfrac{1}{2}\sum\limits_{i\neq j}v(\vert\bs{r}_i-\bs{r}_j\vert,d) \nonumber \\
&- \tilde{V}_m\sum\limits_{i=1}^N\Lambda(\bs{r}_i) + b.g.,
\end{align}
where $v\left(|\mathbf{r}_i - \mathbf{r}_j|, d\right)$ is the electron-electron interaction screened by dual metallic gates, separated by a distance $d$.
The moir\'e potential is characterized by its effective depth $\tilde{V_m} = V_m~\dfrac{m^*/m_e}{(\epsilon/\epsilon_0)^2}$ and shape $\Lambda(\bs{r})$, which
we describe in further detail below. 
The negative sign in front of $V_m$ maps a system of electrons to a system of holes without having to invert the kinetic energy term.
The final term
in Eq.~(\ref{eq:H-2deg-moire})
accounts for a uniform background charge that ensures overall charge neutrality.
 
The presence of metallic gates enables tunability of both the range and strength of the Coulomb interaction via the gate separation, as illustrated in the right panel of Fig.~\ref{fig:schematic}. In momentum space, the dual-gate screened Coulomb interaction admits a closed-form expression~\cite{Yang_PRB2025}
\begin{equation}
v(k, d) = \frac{2\pi \tanh(kd)}{k},
\end{equation}
where $k$ is the magnitude of the wavevector and $d$ is the gate separation. The $\tanh(kd)$ factor suppresses long-range interactions, effectively truncating the Coulomb tail when $d$ is small, allowing the interaction to be tuned from long-ranged to short-ranged.

The presence of the moir\'e potential can effectively enhance the electron-electron interaction. As the moir\'e lattice constant and potential depth increase (at fixed electron density), electrons become more localized at the potential minima, thereby amplifying interaction effects. The shape of the moir\'e potential can be written as a Fourier expansion on the reciprocal space lattice of the moir\'e supercell. The lowest order (harmonic) approximation reads
\begin{equation}\label{eq:potential}
\Lambda(\bs{r}) = \sum_{j=1}^{N/2} 2\cos(\bs{r} \cdot \bs{g}_j + \phi),
\end{equation}
where ${\mathbf{g}_j}$ are the $N$ smallest reciprocal lattice vectors of the moir\'e superlattice. 
In the triangular moiré potential, the phase parameter $\phi$ changes 
the number of local minima of moir\'e potential from one to two and back every $120^\circ$. For the square moir\'e potential, $\phi$ only results in a translational shift of the potential, and we set $\phi=0$.
Examples of moir\'e potentials with different geometries are shown in the middle panel of Fig.\ref{fig:schematic}.

The interaction strength also depends on the electron density. To make this dependence explicit, we can use the Wigner-seitz radius $r_s$ of a homogeneous 2D EG 
as length units, defining $\Tilde{\bs{r}}=\bs{r}/r_s$. For simplicity, we express $r_s$ in units of the effective $a_B^*$. In $r_s$ units, the Hamiltonian reads
\begin{align}\label{eq:effective_Bohr}
\nonumber\hat{H}=&-\frac{1}{2r_s^{2}}\sum_i\tilde{\nabla}_i^2+\frac{1}{2r_s}\sum\limits_{i\neq j}v(\abs{\tilde{\mathbf{r}}_i-\tilde{\mathbf{r}}_j},d)\\
&-\tilde{V}_m\sum_i\Lambda(\tilde{\mathbf{r}}_i)+\mathrm{b.g.},
\end{align}
which explicitly shows the scaling of the kinetic and interaction terms with $r_s$. We use the prefactor of the kinetic term as a proxy to the band width: $W=1/r_s^2$. The electron-electron interaction dominates at low densities (large $r_s$).

\section{Method}\label{sec:method}
Numerical methods play a vital role in complementing ongoing experimental efforts. 
Direct detection of magnetic texture remains an experimental challenge. 
Numerical simulations can readily access spatially resolved magnetic properties, making them especially valuable for exploring and identifying interesting regions of parameter space for magnetic applications.

We apply two separate approaches which complement each other. 
In the first, we perform LSDA calculations on the moir\'e continuum Hamiltonian of Eq.~(\ref{eq:effective_Bohr}). The resulting ground-state properties are presented as a function of the parameters and compared with known properties in the Hubbard model.
In the second approach, we apply Wannierization and downfolding to the continuum Hamiltonian to obtain a lattice model, and examine the resulting model parameters. The results from the two approaches provide cross-checks, which also lead to additional insights on the behavior of the gated 2D EG in the context of model engineering. 
In this section, we introduce our two  computational approaches, describing the DFT with screened interactions in Sec.~\ref{subsec:dft_screened}, and 
the Wannierization and downfolding on top of the DFT results in Sec.~\ref{subsec:wan_downfold}.

\subsection{Density functional theory with screened interactions}\label{subsec:dft_screened}
To study the gated electron gas in a moir\'e potential, we apply DFT with screened interactions.
Practical DFT calculations rely on approximate forms such as 
LSDA.
The conventional form of the exchange-correlation functional $E_{xc}[n,p]$ is parametrized in terms of the local electron density $n$ and spin polarization $p$ via fits to accurate 
total energies obtained from quantum Monte Carlo (QMC) calculations ~\cite{Ceperley_PRL1980, Perdew_PRB1992, Attaccalite_PRL2002}.
However, this form does not account for metallic gate screening, which is essential for modeling gated moir\'e systems with tunable short-range interactions. To overcome this limitation, we adopt the recently developed exchange-correlation functional $E_{xc}[n,p,d]$ parameterized from QMC calculations of  dual-gate screened homogeneous 2D EG, which incorporates the gate separation $d$ as an additional variable~\cite{Yang_PRB2025}.
This functional was implemented within the framework of the libxc library~\cite{Lehtola_SoftwareX2018}.
All DFT calculations are carried out using a modified version of Quantum Espresso adapted for 2D systems~\cite{Giannozzi_JPCM2009,Giannozzi_JPCM2017}.

To determine the ground state, we use simulation cells with various geometries and initialize each self-consistent field (SCF) run with approximately $100$ randomly generated initial configurations. In each configuration, electrons are placed in localized $s$-orbitals centered at random positions. For collinear spin calculations, each electron is assigned a spin orientation (up or down) with equal probability. In addition to collinear spin calculations, we also perform noncollinear spin DFT simulations using the generalized LSDA formalism introduced in Ref.\cite{Kubler_JPFMP1988}.
Most of the noncollinear calculations were initialized with the spin polarization vectors randomly oriented within the 2D plane, but we have checked robustness in select cases with other initial states.
After all SCF runs, the ordered state with the lowest total energy among the converged solutions is selected as the ground state. The kinetic energy cutoff is chosen based on density ($270/r_s$ Ry). To ensure consistent Brillouin zone sampling across different system sizes, we fix the product of the number of $k$-points and the number of unit cells along each lattice direction to 32. Additionally, a $[0.5, 0.5]$ shift is applied to the $\Gamma$-centered Monkhorst-Pack grid to improve sampling. We also performed calculations constrained to the paramagnetic phase using a $4 \times 4$ supercell for Wannierization and downfolding (see Sec.~\ref{subsec:wan_downfold}). The plane-wave kinetic energy cutoff is chosen based on density ($100/r_s$ Ry), and the Brillouin zone is sampled using a $6 \times 6$ $\Gamma$-centered Monkhorst-Pack grid. The convergence threshold for total energy is set to $0.05$ mHa in all DFT calculations.

In previous studies, the Hartree–Fock (HF) method has been widely applied to moir\'e systems ~\cite{Hu_PRB2021,Pan_PRB2020}. Compared to HF, DFT with LSDA offers significantly greater computational efficiency, particularly for large simulation cells.
Furthermore, 
DFT/LDA often yields more accurate results in practice, as HF tends to overemphasize symmetry-broken, ordered states;
an example in 2D EG   moir\'e systems is given recently~\cite{Yang_PRL2024}.
We expect
the accuracy of DFT to sustain, possibly even improve, in moir\'e systems as the gate screening becomes stronger, which 
reduces  
the influence of long-range interactions and thereby makes the system more amenable to local approximations such as LDA.

\subsection{Wannierization and downfolding}\label{subsec:wan_downfold}
To gain insights and facilitate comparison with effective lattice models, we employ two standard post-processing steps: 
Wannierization and downfolding. Wannierization transforms the extended Bloch states into localized real-space orbitals, while downfolding systematically reduces the full Hilbert space to a low-energy subspace that retains the essential physics of interest. Together, these procedures enable a more transparent interpretation of the electronic structure in terms of effective lattice degrees of freedom.

During the Wannierization process, the Kohn–Sham orbitals, initially defined in momentum space and labeled by Bloch wave vectors, are transformed into real-space localized orbitals. 
The spatial localization or spread of the resulting Wannier functions can be tuned by applying a unitary transformation to the set of Kohn–Sham orbitals at each $\mathbf{k}$-point. This unitary transformation is determined by minimizing the total spread of the Wannier functions, following the procedure introduced in Ref.\cite{Marzari_PRB1997} and implemented in the Wannier90 code\cite{Pizzi_JPCM2020}. The resulting maximally localized Wannier orbitals  
serve as a natural basis for constructing effective lattice models. 

In downfolding~\cite{Aryasetiawan_book2022,Chang_NPJCM2024}, we map the 2D EG in a moiré potential to a lattice model. In the DFT band structure for the paramagnetic phase, we identify a set of isolated energy bands that are well separated from higher-energy bands. The number of these isolated bands equals the number of moiré unit cells in the calculation, $N$. 
These $N$ Kohn Sham orbitals are Wannierized following the procedure described above. In this way, we obtain a lattice model in the subspace of the original Hilbert space, where the number of lattice sites equals the number of moiré unit cells in the original model. 

The one-body elements in the resulting low-energy Hamiltonian are obtained from the associated DFT-based tight-binding model in the local Wannier basis.
To incorporate dynamic screening effects, we employ the constrained random phase approximation (cRPA). In this approach, the bare Coulomb interaction within the low-energy (active) subspace is retained, while screening contributions from the remaining (virtual) subspace are computed explicitly at the RPA level. We perform DFT calculations that include a large number of unoccupied bands to evaluate the RPA frequency-dependent dielectric function, and extract the effective interaction parameters by taking the zero-frequency limit~\cite{Yeh_JCTC2023,Yeh_JCTC2024,Aryasetiawan_book2022}. The cRPA downfolding approach improves the resulting Hamiltonian over that from the bare matrix elements, and results in a downfolded lattice model that retains the essential low-energy physics of the original 2D electron system in a moiré potential.

\section{Results}\label{sec:results}

In this section, we present our results, organized into three parts. 
We study the ground state orders of the system described in 
Sec.~\ref{sec:model}, beginning with the square lattice geometry at small gate separation. We then explore the effects of increasing gate separation, followed by a reduction in moiré potential depth, and finally consider the triangular geometry to establish connections with more commonly studied experimental setups.
We compare the results of our continuum Hamiltonian calculations with results from the Hubbard model. In addition, 
we perform
systematic downfolding and Wannierization to obtain model parameters from the original continuum system, to help validate and gauge the connection with the Hubbard model. 

In Sec.~\ref{sec:results-SqSmalld} below, we show that standard square lattice Hubbard model with near-neighbor hopping and on-site interaction can be realized at very small $d$. Note that the statement that $d$ is “small’’ is meant in a relative sense, since our length unit is the effective Bohr radius defined in Sec.~\ref{sec:model}, where $m^*$ and $\epsilon$ are both set to unity for computational convenience. The actual values of these parameters depend on the specific material realization in an experimental setup.
We examine both half-filling and $1/8$-doping using DFT/LSDA with our parametrized functional, and 
show that the ground state of this system reproduces key features of the corresponding ground state 
in the Hubbard model, 
such as antiferromagnetism and stripe phases.
Through the phase diagram of the continuum model and the downfolding analysis, we examine deviations from the
simple Hubbard model as
$d$ increases and the electron 
interaction becomes longer-ranged, as we investigate in Sec.~\ref{sec:results-SqLarged}.  
Our calculations reveal metallic ground states, with  ferromagnetism in a deep moiré potential and noncollinear antiferromagnetic (AFM) order in a shallow moiré potential. We again obtain via downfolding the model parameters, which correspond to the extended Hubbard models.
Finally in Sec.~\ref{sec:results-Tri} we apply the same methodologies to the triangular geometry. We establish a quantitative mapping from current experiments on triangular moiré potentials to the corresponding lattice models. We also summarize, in Table~\ref{tab:proposed_systems}, our downfolding results for some of the moiré systems that have been proposed for realizing the Hubbard model.

\begin{table*}[t]
\caption{\label{tab:proposed_systems}
Downfolding results for some of the candidate moiré systems for Hubbard model realization. For the $\mathrm{MoSe_2/MoS_2}$ system, $m^*$ is taken as the average along the $\mathrm{K-\Gamma}$ and $K-M$ directions for the $C7$ stacking configuration studied in Ref.~\cite{Ji_JCP2018}.
For $\mathrm{MoSe_2/WS_2}$, $m^*$ is taken as the average along the $x$ and $y$ directions in the $AC$ stacking in Ref.~\cite{Li_JCIS2023}. The $V_m$ and $\phi$ for both systems are taken from the $AA$ stacking in Ref.~\cite{Zhang_PRB2020}. In the $U_1/U_0$ column, $\sim 0$ means smaller than our numerical resolution.}
\begin{ruledtabular}
\begin{tabular}{cccccccccccc}
\textrm{Systems}&
\textrm{$\theta$ ($\degree$)}&
\textrm{$a_M$ ($\mathrm{nm}$)}&
\textrm{$V_m$ ($\mathrm{meV}$)}&
\textrm{$\phi$ ($\degree$)}&
\textrm{$m^*$}&
\textrm{$\epsilon$}&
\textrm{$d$ ($\mathrm{nm}$)}&
\textrm{$U_0/t_1$}&
\textrm{$U_1/U_0$}&
\textrm{$t_{\sqrt{3}}/t_1$}&
\textrm{$U_0$ ($\mathrm{meV}$)}
\\
\colrule
$\mathrm{WSe_2/MoS_2}$~\cite{Wu_PRL2018}  & $0$ & $8.5$ & $5.1$  & $-71$ & $0.35$  & $4.2$ & $20.0$ & $26.4$ & $0.50$ & $0.23$ &$55.7$\\
$\mathrm{WSe_2/MoS_2}$~\cite{Wu_PRL2018}  & $0$ & $8.5$ & $5.1$  & $-71$ & $0.35$  & $4.2$ & $5.0$ & $18.1$ & $0.35$ & $0.23$ &$38.1$\\
$\mathrm{WSe_2/MoS_2}$~\cite{Wu_PRL2018}  & $0$ & $8.5$ & $5.1$  & $-71$ & $0.35$  & $4.2$ & $2.0$ & $11.4$ & $0.24$ & $0.23$ &$24.7$\\
$\mathrm{WSe_2/MoS_2}$~\cite{Wu_PRL2018}  & $0$ & $8.5$ & $5.1$  & $-71$ & $0.35$  & $4.2$ & $1.1$ & $8.0$ & $0.18$ & $0.24$ &$16.9$\\
$\mathrm{WSe_2/MoS_2}$~\cite{Wu_PRL2018}  & $0$ & $8.5$ & $5.1$  & $-71$ & $0.35$  & $4.2$ & $0.8$ & $6.4$ & $0.15$ & $0.24$ & $14.2$\\
$\mathrm{WSe_2/MoS_2}$~\cite{Wu_PRL2018} & $0$ & $8.5$ & $5.1$  & $-71$ & $0.35$  & $3.0$ & $0.8$ & $7.9$ & $0.18$ & $0.24$ &$17.5$\\
$\mathrm{WSe_2/MoS_2}$~\cite{Zhang_PRB2020} & $0$ & $8.5$  & $11$  & $40$ & $0.35$  & $4.2$ & $0.8$ & $15.2$ &$0.06$ & $0.17$ &$25.2$\\
$\mathrm{WSe_2/MoS_2}$~\cite{Zhang_PRB2020} & $0$ & $8.5$  & $11$  & $40$ & $0.35$  & $3.0$ & $0.8$ & $18.7$ &$\sim 0$ & $0.17$ &$31.3$\\
$\mathrm{MoSe_2/MoS_2}$~\cite{Zhang_PRB2020,Ji_JCP2018} & $0$ & $8.3$ & $9$  & $42$ & $0.38$  & $4.2$ & $0.8$ & $12.3$ &$\sim 0$ & $0.19$ &$21.9$\\
$\mathrm{MoSe_2/MoS_2}$~\cite{Zhang_PRB2020,Ji_JCP2018} & $0$ & $8.3$  & $9$  & $42$ & $0.38$  & $3.0$ & $0.8$ & $15.1$ &$\sim 0
$ & $0.19$ &$27.0$\\
$\mathrm{MoSe_2/WS_2}$~\cite{Zhang_PRB2020,Li_JCIS2023} & $0$ & $8.2$ & $7$  & $35$ & $0.64$  & $4.2$ & $0.8$ & $28.4$ &$\sim 0$ & $0.15$ & $25.4$\\
$\mathrm{MoSe_2/WS_2}$~\cite{Zhang_PRB2020,Li_JCIS2023} & $0$ & $8.2$  & $7$  & $35$ & $0.64$  & $3.0$ & $0.8$ & $33.7$ &$\sim 0$ & $0.15$ & $30.1$\\
$\mathrm{WSe_2/MoSe_2}$~\cite{Wu_PRL2018} & $1$ & $19$ & $6.6$  & $-94$ & $0.35$  & $4.2$ & $0.8$ & $352$ &$\sim 0$ & $0.01$ & $16.0$\\
\end{tabular}
\end{ruledtabular}
\end{table*}

\subsection{Realization of the Hubbard model}\label{sec:results-SqSmalld}
We study the 2D EG 
in a square moiré potential at both half-filling ($\nu=1$) and $1/8$ hole doping ($\nu=7/8$).  
In the conventional Hubbard model, half-filling corresponds to one electron per lattice site, with each site hosting a single orbital that can accommodate up to two electrons with opposite spins. By contrast, in our moiré system, each moiré potential valley contains multiple energy levels. We use $\nu=1\; (7/8)$ to represent the case where each moir\'e unit cell is occupied by $1\; (7/8)$ electrons on average.

At half-filling, the Hubbard model is known to exhibit a Mott insulating ground state with checkerboard AFM for any finite interaction strength 
$U/t$. 
Away from half-filling, however, the model exhibits increased complexity and lacks an exact solution. Recent numerical studies~\cite{Zheng_Science2017} indicate that the stripe phase is the ground state at $1/8$ hole doping, 
when $U/t$ 
exceeds a critical threshold~\cite{Xu_PRR2022}.

Below we discuss the two fillings in two subsections. 
We observe in the 2D EG system both the checkerboard AFM phase and the stripe phase. 
The properties of these phases are consistent 
with state-of-the-art results from the Hubbard model.
Moreover, the parameter values obtained through downfolding 
are consistent with a simple Hubbard model, with hopping amplitudes $t_n$ for $n>1$ (beyond near-neighbor) and interaction strength $U_n$ for $n>0$ (beyond on-site) both strongly suppressed. 
In the stripe phase, 
the critical ratio $U_0/t_1$ obtained through downfolding agrees quantitatively with that of the square Hubbard model, reinforcing the validity of our moiré-engineered realization.

\subsubsection{Half-filling ($\nu=1$)}
\begin{figure}[h]
\includegraphics[trim=0cm 0.8cm 0cm 0.6cm, clip, width=0.5\textwidth]{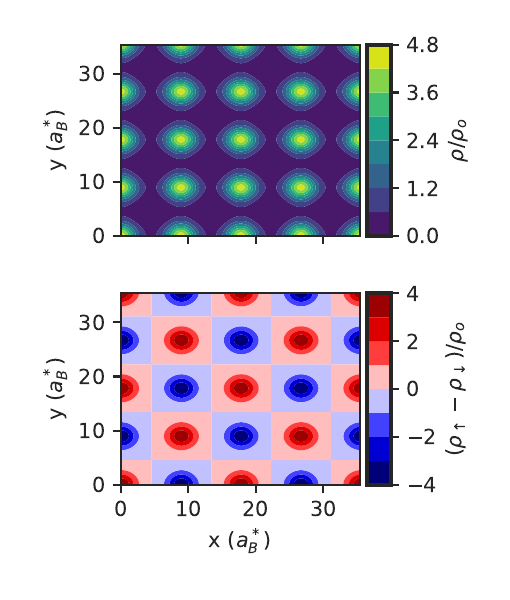}
\centering
\caption{Charge and spin patterns in the ground state of system $\nu=1$, $r_s=5\; a_B^*$, $V_m/W=1.6$, and $d=0.3\; a_B^*$. The upper panel is the charge density and the lower panel the spin density. Both densities are normalized by the average charge density $\rho_0=N_e/A$, where $N_e$ is the number of electrons and $A$ the area of the simulation cell.}
\label{fig:checkerboard}
\end{figure}

For $\nu = 1$, we systematically explore the parameter space defined by $0.1 \leq d \leq 0.9$ (in units of $a_B^*$) and $1.1 \leq V_m/W \leq 2.0$. 
We find  
that the insulating 
AFM phase is robust 
in most of this region. An example of the corresponding charge and spin configuration is shown in Fig.~\ref{fig:checkerboard}. This result is consistent with the half-filled Hubbard model, whose ground state is known to be a Mott insulator with AFM order for any finite $U/t$.

However, a notable difference arises in the moiré system: within a narrow window at small gate separation ($d = 0.1\; a_B^*$) and moiré potential strength ($V_m/W = 1.1$–$1.3$), we observe a paramagnetic metallic phase. Our DFT/LSDA results indicate that the transition from the AFM to the paramagnetic phase coincides with an insulator-to-metal transition.

We quantitatively connect the continuum moiré model to an effective lattice model through the downfolding procedure described in Sec.~\ref{subsec:wan_downfold}. The downfolded parameters reveal that the critical effective interaction strength $U_0/t_1$ for the AFM insulator to the paramagnetic metal phase transition is between $2.2$ and $2.5$. Notably, we also find that the diagonal (next-nearest-neighbor) hopping $t_{\sqrt{2}}$ is significantly suppressed compared to the third-nearest-neighbor hopping $t_2$. For instance, at $V_m/W=1.6$ and $d=0.3\;a_B^*$, $t_{\sqrt{2}}/t\approx0.003$ while $t_2/t\approx0.08$; at $V_m/W=1.1$ and $d=0.1\;a_B^*$, $t_{\sqrt{2}}/t\approx0.001$ while $t_2/t\approx0.11$. This suppression arises from the presence of potential maxima located along the diagonals between neighboring potential minima, which effectively inhibit electron tunneling along these paths.

\subsubsection{$1/8$-doping ($\nu=7/8$)}
\label{ssec:Hub-doped}
\begin{figure}[h]
\includegraphics[width=0.5\textwidth]{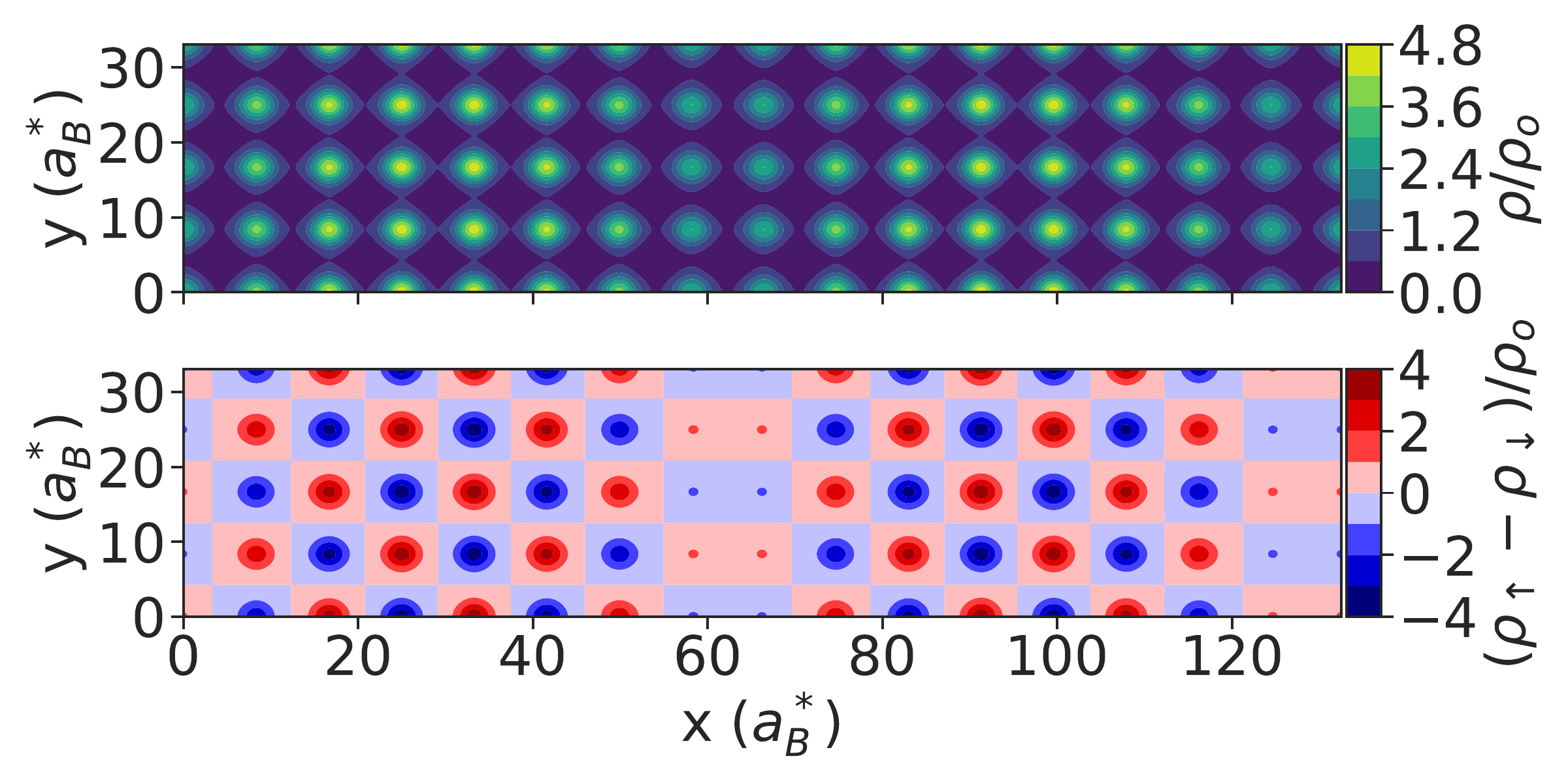}
\centering
\caption{Charge and spin patterns in the ground state of system $\nu=7/8$, $r_s=5\; a_B^*$, $V_m/W=1.6$, and $d=0.3\; a_B^*$. Setting and conventions are similar to Fig.~\ref{fig:checkerboard}.}
\label{fig:spinNcharge}
\end{figure}

We study our square moiré system at filling $\nu=7/8$, i.e., $1/8$-hole-doped in the convention of the Hubbard model. 
The phase diagram scanning a range of gate separation $d$
and moiré potential depth is shown in Fig.~\ref{fig:phaseDiagram}. In this section, we focus on small $d$ and deep potential, 
in particular $V_m/W=1.6$.
We find a stripe phase at  $d=0.3\,a_B^*$  as illustrated in Fig.~\ref{fig:spinNcharge}. 
In the charge density plot, we identify periodically spaced regions of suppressed charge density. 
Simultaneously, the spin density shows checkerboard AFM order, interrupted by domain walls that lie 
within the stripe regions. These domain walls introduce a $\pi$ phase shift in the AFM order.
This state is consistent with the stripe phase found in the Hubbard model at zero temperature~\cite{Zheng_Science2017,Xu_PRR2022}. Following the literature, we refer to this as a linear stripe phase\cite{Xu_JPCM2011, Zheng_Science2017}. Within each stripe period along a row, there is approximately one hole on average, characterizing this configuration as a filled stripe\cite{Zheng_Science2017}.

\begin{figure}[h]
\includegraphics[width=0.5\textwidth]{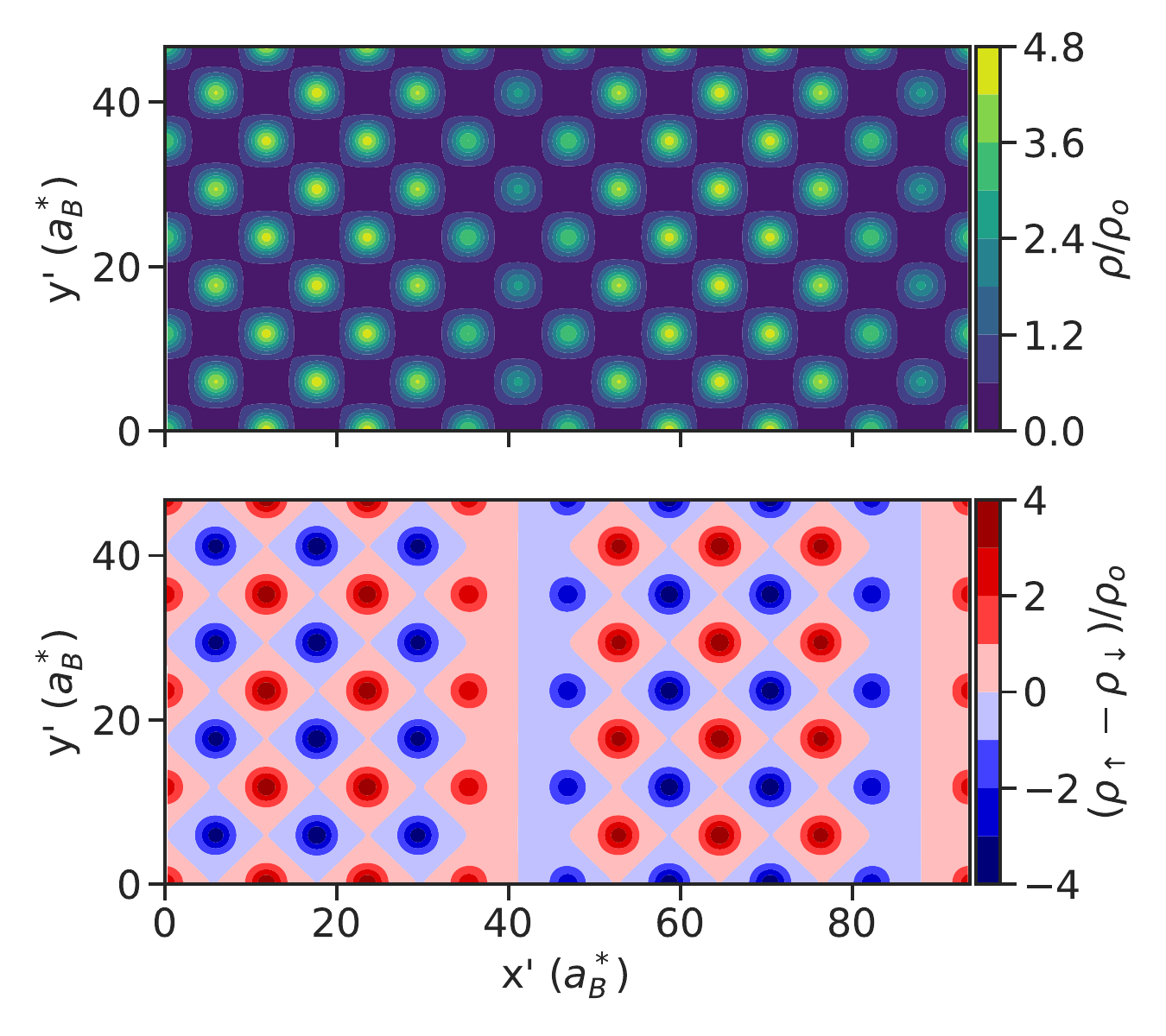}
\centering
\caption{Diagonal charge and spin patterns in the ground state of system $V_m/W=1.6$ and $d=0.4\; a_B^*$. 
In this plot, the horizontal and vertical axes, denoted by $x'$ and $y'$, 
represent the $(1,1)$- and $(-1,1)$-directions, respectively. 
That is, they are rotated by $45\degree$ with respect to the  $x$- and $y$-axes in Figs.~\ref{fig:checkerboard}
and \ref{fig:spinNcharge}; other conventions are the same. 
}
\label{fig:spinNcharge_diag}
\end{figure}

As the gate distance $d$ increases, the system undergoes a transition to a diagonal stripe phase, where both charge and spin modulate along the [$1$, $1$] direction (see Fig.\ref{fig:spinNcharge_diag}).
The modulation patterns are similar to the linear stripe case, but the stripe orientation rotates by $45$ degrees. Additionally, the charge density minimum shifts from bond-centered (in the linear stripe) to site-centered (in the diagonal stripe). This is also manifested in the positions of the domain walls in the spin density plot. However, since the stripe pattern is spatially extended over several sites, the distinction between bond-node and site-node stripes becomes less critical~\cite{Zheng_Science2017}.

When mapped to  a lattice model through Wannierisation and downfolding, 
we find that these parameter values of $d$ and $V_m/W$ correspond to 
$U/t$ in the range of $4$ to $8$,
with essentially no interaction beyond on-site and only 
small hopping amplitude $t_n/t_1$ beyond near-neighbor (further details in the next section). The ground-state orders seen above are compatible with the stripe phase found in the Hubbard model at zero temperature~\cite{Zheng_Science2017,Xu_PRR2022}. 
An example of a quantitative comparison is given below.
By scanning $d$,
we find the range of the critical value $U/t$ for the onset 
of the stripe order in agreement with
the phase diagram of the Hubbard model (no next-near-neighbor hopping) which places the stripe phase boundary at 1/8-doping 
between $U/t = 3.5$ and $5$
~\cite{Xu_PRR2022}. 
Moreover, we scan the range $1.1\le V_m/W\le 2.0$ and find that such an agreement holds for $V_m/W\geq1.6$. The disagreement for $V_m/W<1.6$ is likely due to the increasing influence of further neighbor hoppings.

\begin{figure}[h]
\includegraphics[width=0.5\textwidth]{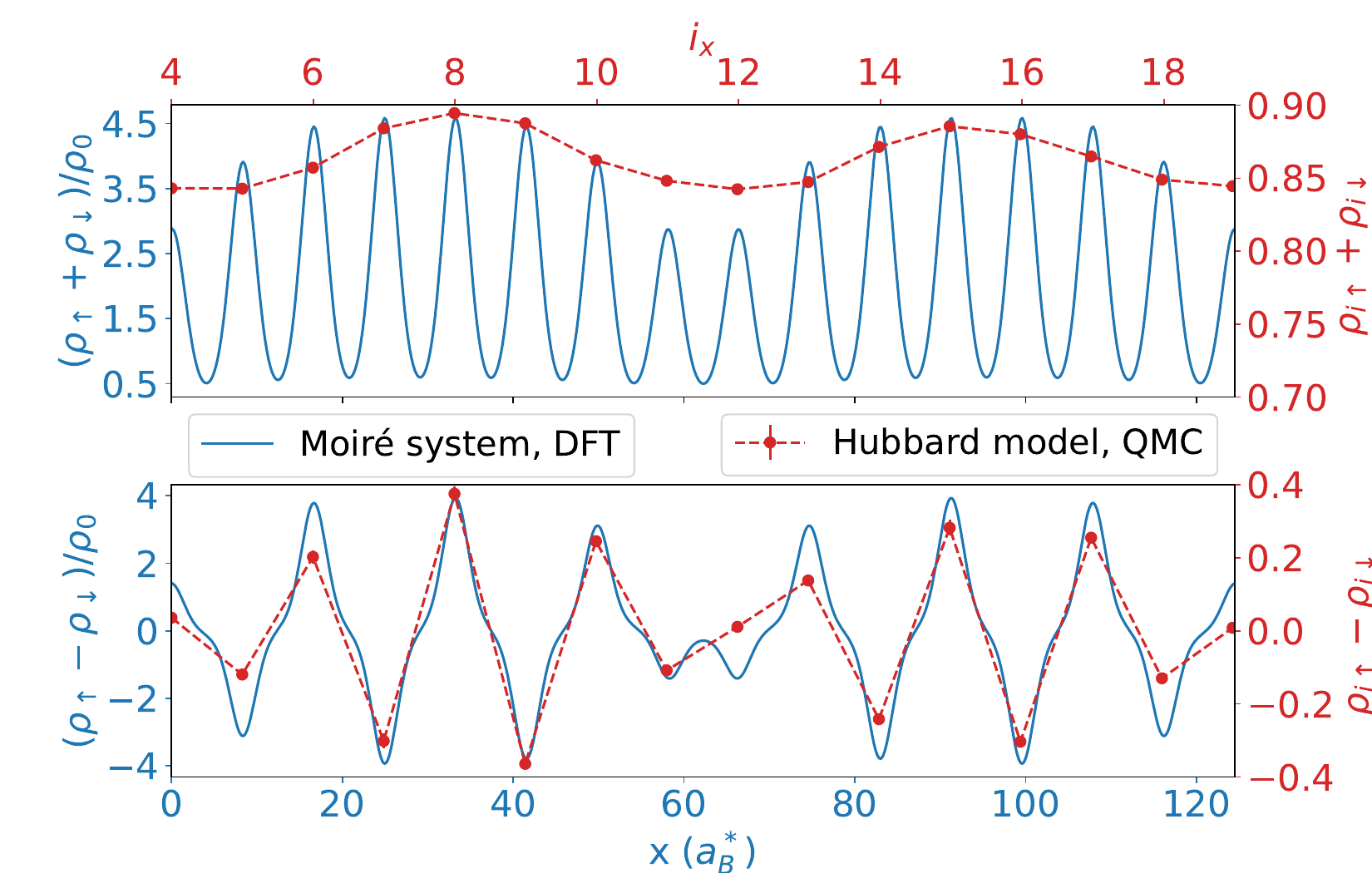}
\centering
\caption{Charge and spin density line cuts in the ground state of system $\nu=7/8$, $r_s=5\; a_B^*$, $V_m/W=1.6$, $d=0.4\; a_B^*$. The upper panel is the charge density and the lower panel the spin density. The blue curves are DFT results for the moiré system. For comparison, the red curves are results from CP AFQMC~\cite{Zhang_PRL1995,Zhang_PRB1997} calculations in the square-lattice Hubbard model on a ($L_x$, $L_y$)=($4$, $32$) at $U/t=8$
($i_x$ and $i_y$ label the lattice sites along the $x$ and $y$ directions.) Error bars are smaller than the marker size. In both panels, the Hubbard model sites are aligned with the minima of the moiré potential. The slight asymmetry of the red curves arises from the effect of the pinning fields in the finite system.}
\label{fig:dft_vs_qmc}
\end{figure}

As an example, we show in Fig.~\ref{fig:dft_vs_qmc} a comparison between the DFT result in our moiré system and  results in the Hubbard model from constrained-path auxiliary field quantum Monte Carlo (CP AFQMC)~\cite{Zhang_PRL1995,Zhang_PRB1997} calculations.
The CP AFQMC results are obtained following the same procedures as in Refs.~\cite{Qin_PRB2016,Zheng_Science2017,Xu_PRR2022}. Open (periodic) boundary condition is applied in $y$ ($x$) direction, and pinning fields ($v_p=0.1$) are applied on the two edges of the cylinder. The trial wavefunction is the unrestricted HF solution to the same model with an effective onsite interaction $U_{\mathrm{eff}}=2.7$, determined from a self-consistent approach~\cite{Qin_PRB2016}. 
We find that the envelopes of the charge and spin density in the moiré system closely reproduce those of the Hubbard model when the moiré potential minima are mapped onto the Hubbard sites. The continuum model offers additional details.
Both the charge and spin densities are strongly localized at the potential valleys. Between neighboring valleys, the spin density decreases to nearly zero, whereas the charge density remains finite at approximately $0.5$.

A natural question can arise regarding the reliability of DFT/LSDA, 
 in treating the Hamiltonian in Eq.~(\ref{eq:effective_Bohr}). 
There are two aspects which provide strong support for the validity of our 
conclusions. First, 
since the system is a 2D EG in an external potential which is rather smooth, it is 
reasonable that an LSDA treatment based on a functional specifically parametrized for
the homogeneous 2D EG (with the exact same gate-screened interaction) can capture 
the charge and spin properties. Indeed a direct comparison under the triangular geometry 
between DFT/LSDA and QMC calculations shows that the former 
is rather accurate in predicting the ground-state phases ~\cite{Yang_PRL2024}.
Second, in the Hubbard model it is known that unrestricted Hartree-Fock gives 
reasonable predictions on the stripe order~\cite{Xu_JPCM2011,Xu_PRR2022}, especially 
with some renormalization of the interaction strength $U/t$ \cite{Qin_PRB2016,Xu_PRR2022}. In some sense, our LSDA calculation can be compared with 
the unrestricted HF (UHF) in the Hubbard model.
Interestingly, the diagonal stripe order we 
observe above in Fig.~\ref{fig:spinNcharge_diag} is also found in the Hubbard model at larger $U/t$ in UHF~\cite{Xu_JPCM2011}, while its existence is more ambiguous in more accurate 
QMC calculations~\cite{Xu_PRR2022}.
These observations all indicate 
that the system indeed 
realizes the standard square lattice Hubbard model. 

\subsection{Larger $d$ and the extended Hubbard model}\label{sec:results-SqLarged}

\begin{figure}[h]
\includegraphics[width=0.5\textwidth]{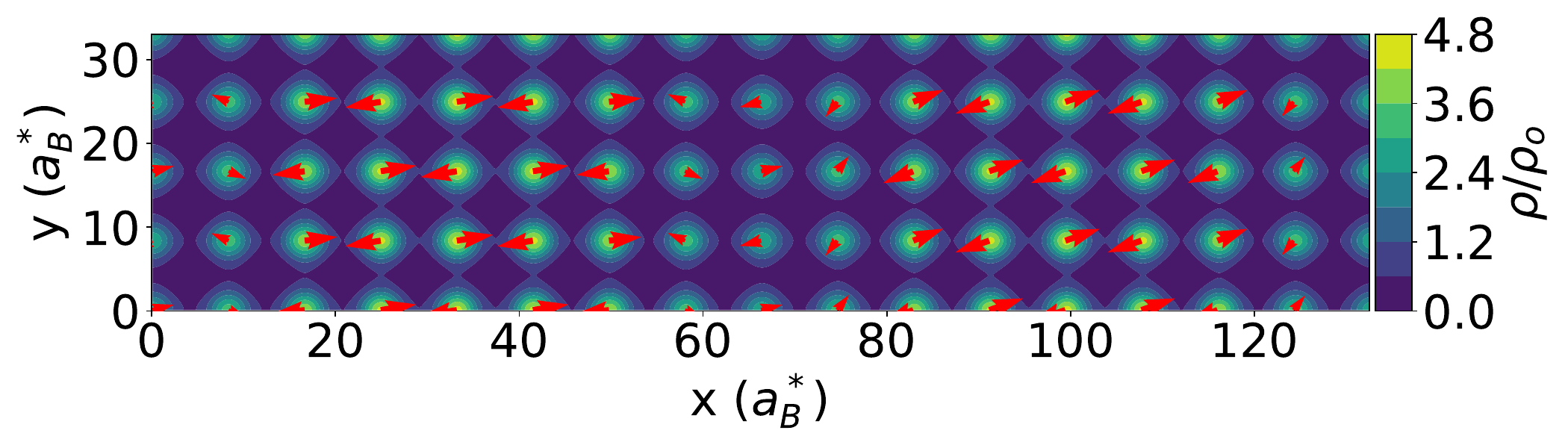}
\centering
\caption{Charge and spin patterns in the ground state of system $\nu=7/8$, $r_s=5\; a_B^*$, $V_m/W=1.5$, $d=0.7\; a_B^*$. The red arrows denote the spin orientations in the $x$-$y$ plane. The maximum out-of-plane ($z$) spin component is comparable to the maximum $y$ spin component.}
\label{fig:noncolinear_stripe}
\end{figure}

The results from the previous section confirm the realization of the square lattice Hubbard model at small $d$. Here we stay at $\nu=7/8$ and scan a larger range of $d$ as well as the moiré potential depth. The phase diagram is shown in
Fig.~\ref{fig:phaseDiagram}.
The two systems discussed in Sec.~\ref{ssec:Hub-doped} fall on the vertical dashed line, where stripe phases are seen, with
charge and spin densities exhibiting modulations along the [$1$, $0$] direction 
at $d=0.3$ and along [$1$, $1$] at $d=0.4\,a_B^\star$. 
At even larger $d$, we observe a transition to a non-colinear stripe phase. For example, at $V_m/W = 1.4$, $d = 0.8\;a_B^*$ and $V_m/W = 1.5$, $d = 0.7\;a_B^*$, the ground state exhibits non-colinear spin textures superimposed on a stripe-like charge distribution, as shown in Fig.~\ref{fig:noncolinear_stripe}.
These states are insulating from our DFT/LSDA band structure. 

We also observe that the stripe phase becomes metallic when the moiré potential 
becomes shallower.
These metallic stripe phases exhibit longer periodicities compared to their insulating counterparts and are marked by purple triangles in Fig.~\ref{fig:phaseDiagram}. 
As the gate distance $d$ decreases, the stripe phase transitions into a metallic checkerboard AFM phase. 
A paramagnetic phase emerges as the interaction strength 
is further reduced. At $d = 0.1\;a_B^*$, decreasing the moiré potential from $V_m/W = 1.6$ to $1.5$ 
causes the ground state to transition from a metallic checkerboard AFM phase to a metallic paramagnetic phase. This is in contrast to the half-filling case studied in Sec.~\ref{sec:results}, where the insulator-to-metal transition is accompanied by a transition from the AFM to the paramagnetic phase.

\begin{figure}[h]
\includegraphics[width=0.5\textwidth]{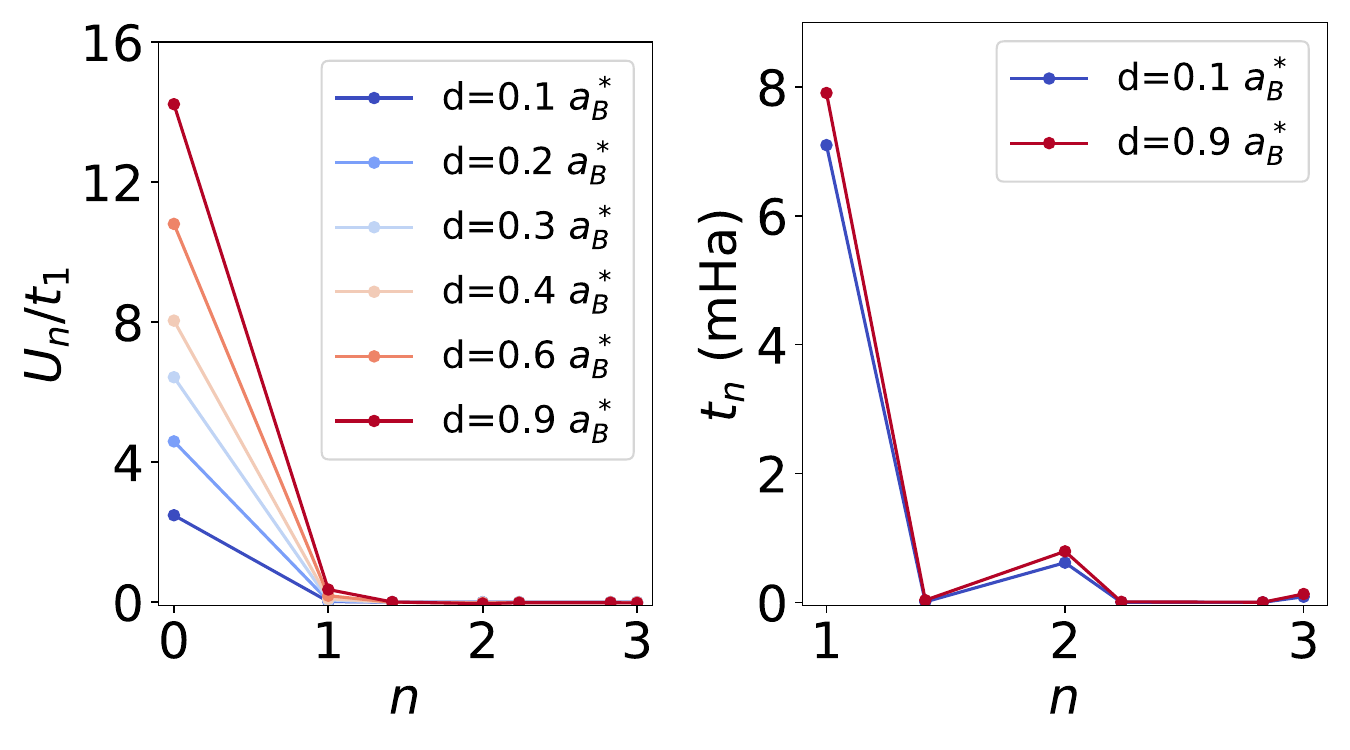}
\centering
\caption{The extended Hubbard model parameters obtained from downfolding at $\nu=7/8$, $r_s=5\; a_B^*$, and $V_m/W=1.6$. The calculations are performed on a $4\cross4$ cell, and follow the procedure described in Sec.~\ref{subsec:wan_downfold} and applied in the $\nu=1$ case. The system parameters are denoted by the vertical dashed black line in Fig.~\ref{fig:phaseDiagram}. The left panel displays the orbital interactions $U_n$ normalized by the nearest neighbor hopping amplitudes $t_1$. The right panel plots the corresponding hopping amplitudes as a function of the orbital separation in the number of lattice constants $n$. The moiré potential peaks suppress the diagonal-related hopping.
}
\label{fig:hubbLineCut}
\end{figure}

\begin{figure*}
\includegraphics[width=1.0\textwidth]{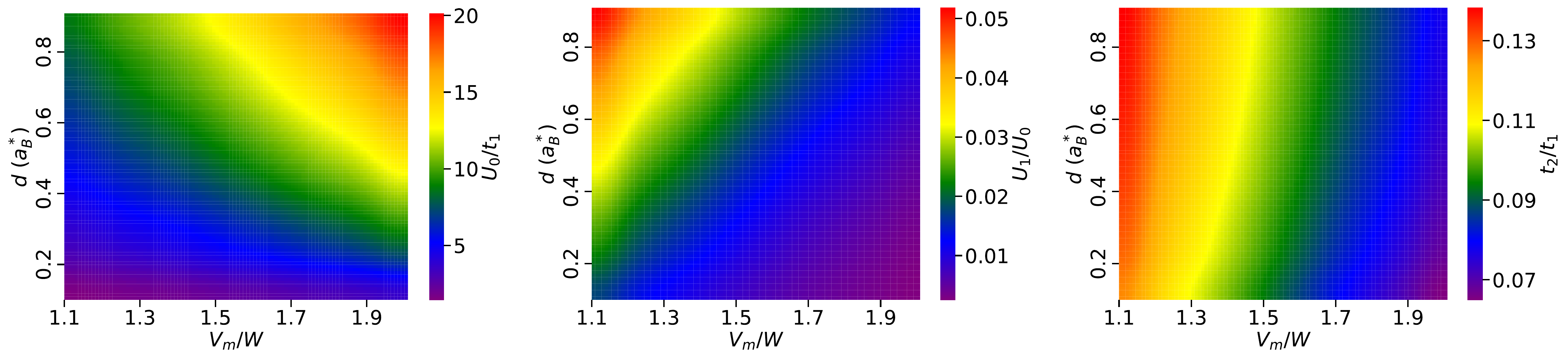}
\centering
\caption{Heat maps of the parameters from downfolding as functions of gate distance $d$ and the moiré potential depth normalized by the kinetic energy $V_m/W$: Intra-orbital interaction $U_0$ normalized by the nearest neighbor hopping amplitude $t_1$ (left), the nearest neighbor interaction $U_1$ normalized by $U_0$ (center), and the third nearest neighbor hopping amplitude $t_2$ normalized by $t_1$ (right).}
\label{fig:hubbHeatPlot}
\end{figure*}

We map the main portion of the parameter space  in 
Fig.~\ref{fig:phaseDiagram}
into an effective lattice model
through Wannierization and downfolding. We follow the same procedure as described 
in Sec.~\ref{subsec:wan_downfold} and applied in the $\nu=1$ case above. 
The downfolded parameters along the vertical line cut in Fig.~\ref{fig:phaseDiagram} are presented in Fig.~\ref{fig:hubbLineCut}.
The results show that hopping amplitudes beyond nearest neighbors are small, and inter-orbital interactions are negligible, as mentioned in 
Sec.~\ref{ssec:Hub-doped}.
The downfolded Hamiltonian 
corresponds to a $t$-$t'$-$t''$-$U$ Hubbard model, with very small $t'$ ($t'/t\sim0.002$) and a $t''/t\sim 0.09$. 
Such a small but non-negligible $t''$ might help stabilize the metallic checkerboard phase for small $U_0/t_1$.
To have a more global view of 
the model parameters in the 
space of $0.1 \leq d \leq 0.9$ and $1.1 \leq V_m/W \leq 2.0$, 
we plot 
the corresponding values of the interaction ratio $U_0/t_1$, the nearest-neighbor interaction ratio $U_1/U_0$, and the next-nearest-neighbor hopping ratio $t_2/t_1$
as heat maps in Fig.~\ref{fig:hubbHeatPlot}.
We see that throughout 
this parameter region, 
the ratio of inter-orbital interactions to $t_1$ and the hopping amplitudes beyond the nearest neighbors are negligible compared to $U_0/t_1$ and $t_1$, respectively, indicating that our moiré system still  
approximates a Hubbard model
reasonably well.

\begin{figure}
\includegraphics[width=0.5\textwidth]{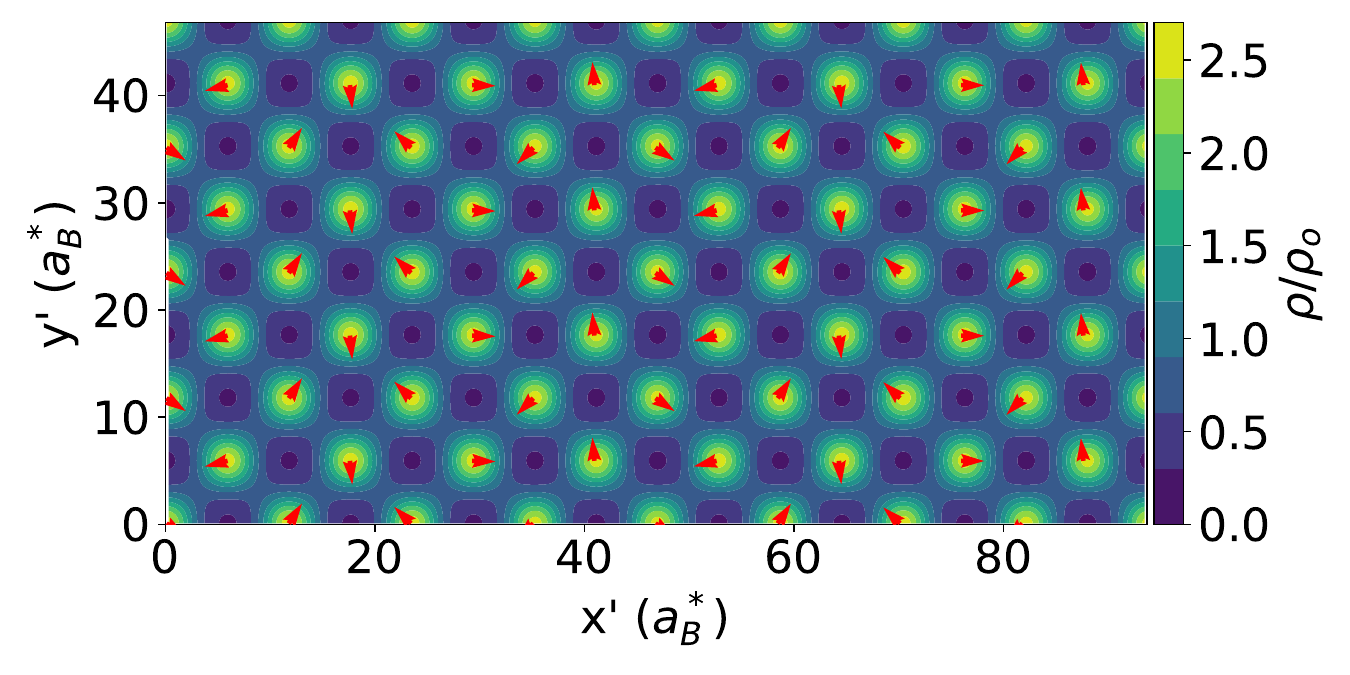}
\centering
\caption{Charge and spin patterns in the ground state of system $\nu=7/8$, $r_s=5\; a_B^*$, $V_m/W=0.7$, and $d=3\; a_B^*$. In this plot the horizontal and vertical axes, denoted by $x'$ and $y'$, 
represent the $(1,1)$- and $(-1,1)$-directions, respectively. 
That is, they are rotated by $45\degree$ with respect to the  $x$- and $y$-axes in Figs.~\ref{fig:noncolinear_stripe}. The red arrows denote the spin orientation in the $y$-$z$ plane. The maximum out-of-plane ($x$) spin component is around $1/5$ of the maximum of the $y-z$ plane component.}
\label{fig:spiral}
\end{figure}

\begin{figure}
\includegraphics[width=0.5\textwidth]{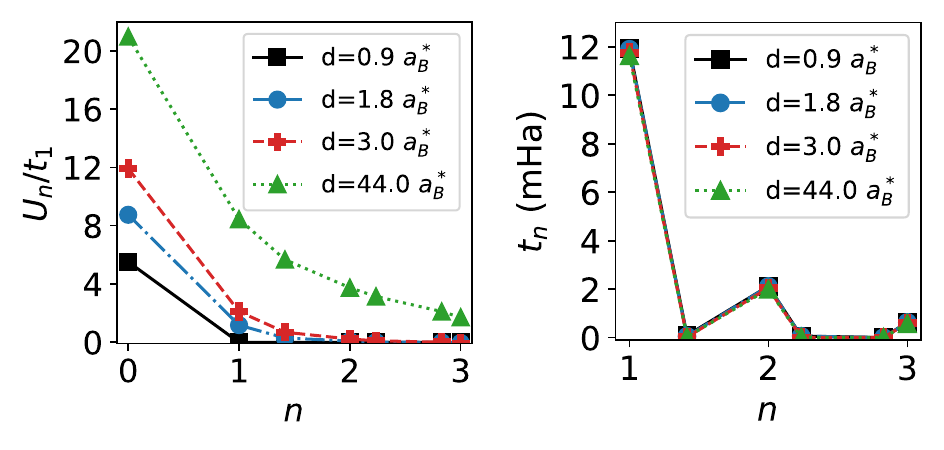}
\centering
\caption{The downfolding parameters for systems at $\nu=7/8$, $r_s=5\; a_B^*$, $V_m/W=0.7$, and 
$d=0.9,\; 1.8,\; 3.0,\;$ and $44.0\; a_B^*$.
The layout is similar to Fig.~\ref{fig:hubbLineCut}.
}
\label{fig:checkerboard2spiral}
\end{figure}

We have also investigated 
two other regimes with even larger gate separation $d$: shallow and deep moiré potentials. 
In the former, 
a shallow moiré potential ($V_m/W=0.7$), 
we observe a transition from a metallic checkerboard AFM phase to a metallic noncollinear AFM phase
as $d$ increases from $1.8\;a_B^*$ to $3\;a_B^*$.
The corresponding charge and spin order are shown in Fig.~\ref{fig:spiral}.
In the noncollinear AFM phase, the charge density retains the discrete translational symmetry of the moiré potential, while the spin order becomes noncolinear with the spin orientations modulating along the [$1$,$1$] direction. The downfolding parameters associated with this transition are plotted in Fig.~\ref{fig:checkerboard2spiral}.
We can see that the hopping amplitudes are not significantly affected by the gate screening, while both the onsite and longer-range interactions increase with $d$,
as expected.

\begin{figure}[h]
\includegraphics[width=0.5\textwidth]{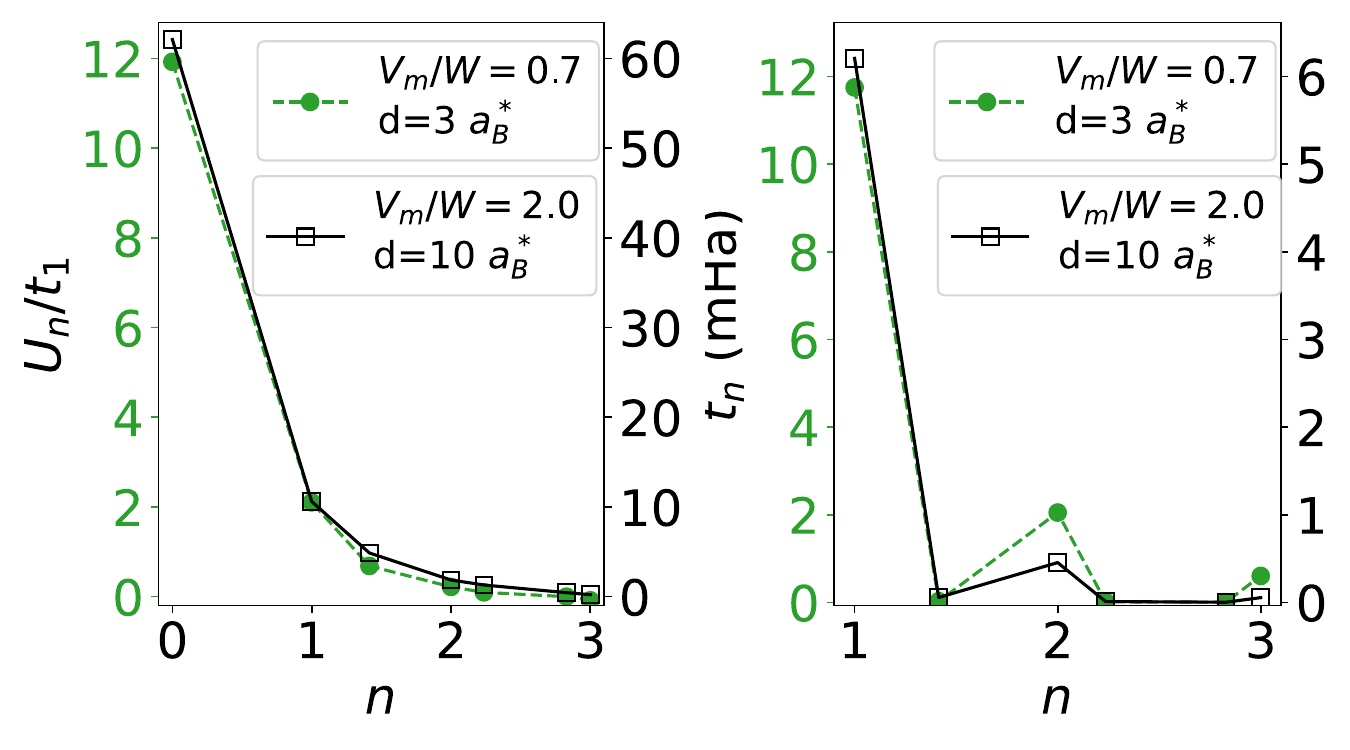}
\centering
\caption{The downfolding parameters for systems $\nu=7/8$, $r_s=5\; a_B^*$, with $V_m/W=0.7$, $d=3\; a_B^*$ and $V_m/W=2.0$, $d=10\; a_B^*$, respectively.
The layout is 
similar to Fig.~\ref{fig:hubbLineCut}.
}
\label{fig:spiralFerroLineCut}
\end{figure}

In the case of large $d$ and
a deep moiré potential, 
the ground state becomes a ferromagnetic metal
at $V_m/W=2.0$, $d=10\, a_B^*$. 
The downfolding parameters for this case are shown in Fig.~\ref{fig:spiralFerroLineCut} 
together with those for the system with $V_m/W=0.7$ and $d=3\,a_B^*$. In the former, $U_0/t_1$ is approximately $5$ times larger. A strong on-site interaction can favor ferromagnetism, as is given by the Stoner criterion~\cite{Scalettar_LectureNotes2016}. Such a phase has been shown to appear in HF calculations on the Hubbard model~\cite{Xu_JPCM2011}. 
As discussed earlier, in the strong interaction regime, it is plausible that 
our LSDA calculation in the 2D EG
with external potential behaves 
in a manner more similar to UHF (than to more accurate many-body results).

\subsection{Triangular lattice}\label{sec:results-Tri}

\begin{figure*}
\includegraphics[width=1.0\textwidth]{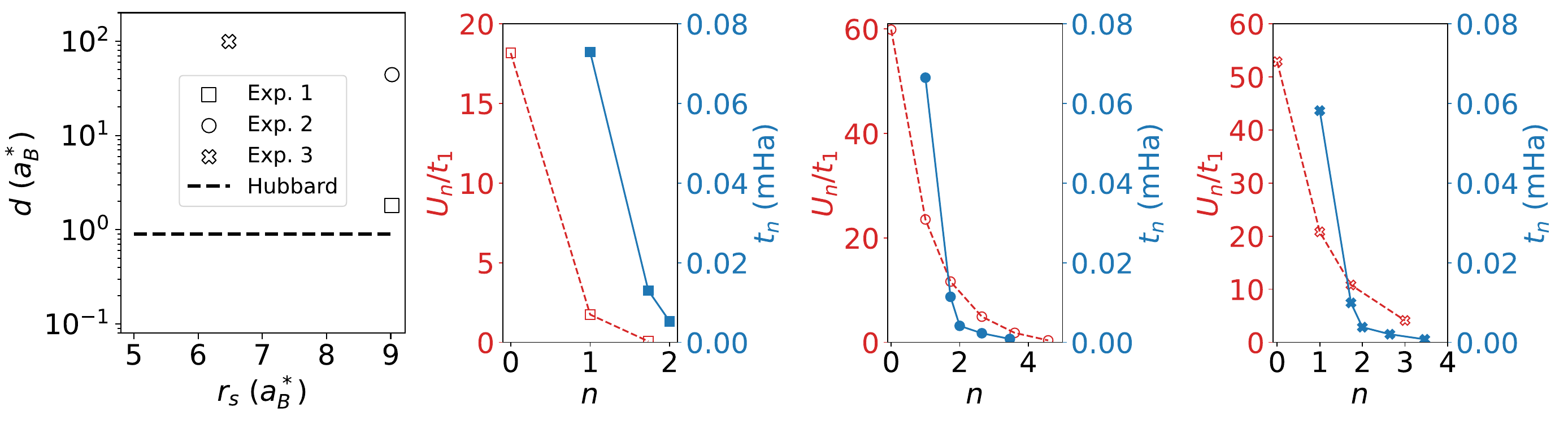}
\centering
\caption{The downfolding parameters for the existing experiments on triangular moiré potential. Hubbard physics appears for $d<0.9 \;a_B^*$ in Fig.~\ref{fig:phaseDiagram}. This region is below the horizontal dashed black line in the first panel. Three sets of experimental parameters on triangular moiré potential are represented by the open markers, where Exp. 1 (second panel) and Exp. 2 (third panel) are taken from Refs.~\cite{Tang_NN2023,Tang_Nature2020}, and Exp. 3 (fourth panel) from Ref.~\cite{Regan_Nature2020}.}
\label{fig:triangular}
\end{figure*}

Experiments in moiré systems have already specifically targeted the 
triangular Hubbard model~\cite{Tang_Nature2020,Regan_Nature2020,Tang_NN2023}. In these experiments, the holes in the valence band top of $\mathrm{WSe_2}$ play the role of electrons (with the effective mass 0.35)
~\cite{Wu_PRL2018}. 
These holes experience a moiré potential arising from the lattice mismatch between the top layer of $\mathrm{WS_2}$ and the bottom layer of $\mathrm{WSe_2}$, with a moiré lattice constant $a_M$ of about $8\;nm$. The interaction between the holes are screened by metallic gates. Hexagonal boron nitride (hBN) layers separate the metallic gates from the moiré system, providing a dielectric constant 
$\epsilon$ of either $3$\cite{Tang_Nature2020,Tang_NN2023} or $4.2$~\cite{Regan_Nature2020}. Following the treatment in Eq.~(\ref{eq:effective_Bohr}), we find this results in an effective Bohr radius $a_B^*$ of approximately $8.6$ and $12.0$, respectively. We summarize three experimental realizations of the triangular Hubbard model in $\mathrm{WS_2}/\mathrm{WSe_2}$ bilayers in Fig.~\ref{fig:triangular}. The variation in $r_s$ (in units of $a_B^*$) arises from differences in the dielectric constant $\epsilon$ of the dielectric layers.

To quantitatively connect the existing experiments on the triangular moiré potential to lattice models, we again apply DFT and downfolding as we have done in the square geometry above.
As a representative example, we consider the 
experiment of Tang \textit{et al.}
in Ref. ~\cite{Tang_NN2023} (referred to as EXP. 1). 
The bottom gate placed approximately $0.8\;nm$ from the 2D system, while the top gate was kept at a distance of $20$ nm. To apply our density functional parametrized for a symmetric-dual-gate-screened system, we considered a system with both gates $0.8\;nm$ away from the 2D system. This corresponds to $1.8\;a_B^*$ when we take $\epsilon\approx3$ for the dielectric constant $\mathrm{hBN}$~\cite{Tang_Nature2020,Tang_NN2023}. Accordingly, $r_s=9.02a_B^*$ at $\nu=1$. We use the moiré potential $V_m=15 \;\mathrm{meV}$ and the parameter $\phi=\ang{45}$ in Eq~(\ref{eq:potential}), as fitted by Zhang et al.~\cite{Zhang_PRB2020} from DFT results assuming relaxed interlayer distance and rigid in-plane lattice. In Hartree atomic units, this corresponds to $V_m=14.2 \;\mathrm{mHa}^*$.

The downfolding results are presented in Fig.~\ref{fig:triangular}. 
Our calculation show that while gate screening does shorten the electron-electron interaction range in EXP. 1, the onsite interaction remains substantial, with $U_0/t_1=18$. 
So this is more consistent with the triangular lattice Hubbard model realized in the strong interaction limit.
In contrast, when the gate separation is larger ($d>20\;nm$), the interactions become long-ranged and the ratio $U_0/t_1$ increases by a factor of $3$. 
These situations are also distinct from the parameter regime we were targeting above, and realize extended Hubbard models in the triangular lattice. 

To quantify how the interaction and hopping parameters of lattice models realized in moiré systems are controlled by experimentally accessible knobs, we perform downfolding for several theoretically proposed systems. We set $d$ to be equal or above the smallest $d$ ($0.8\;\mathrm{nm}$) realized in experiments on $\mathrm{WS_2/WSe_2}$ bilayers (EXP.~1). The results are summarized in Table~\ref{tab:proposed_systems}. We find that moiré-material–dependent properties, such as the moiré lattice constant, moiré potential, and the effective mass of the charge carriers, jointly determine the interaction strengths together with the gate separation and dielectric environment. The moiré potential in realistic devices can be larger than the theoretical values adopted here from the literature~\cite{Shabani_NP2021}. For a given moiré system and a fixed lower bound for $d$, the interactions can be efficiently reduced by increasing the $\epsilon$ of the dielectric environment. In contrast, the hoppings are insensitive to variations in $\epsilon$ and $d$.

\section{Conclusions \& Outlook}\label{sec:Conclusions}

In this paper, we presented a systematic and quantitative study to show how a 2D EG in a moiré potential can serve as a tunable quantum simulator for lattice models, by controlling the potential and the metallic gates. In particular, with a square moiré potential, the system can be made to approach the iconic Hubbard model which has been a hallmark in modern condensed matter physics. We showed that stripe phases appear and show 
properties consistent with those in the ground state of the square lattice Hubbard model. 
Downfolding the moiré system to a lattice model, we find that the parameters are consistent with the Hubbard model which exhibits similar phases. 
As the range and strength of the electron-electron interactions are varied via gate separation and moiré well depth, we observe 
a rich variety of metallic magnetic orders, including checkerboard AFM, noncollinear AFM order, paramagnetism, and ferromagnetism. We obtain the corresponding lattice models via downfolding. 
Comparing with the triangular moiré systems already realized in experiments, weaker and shorter range interaction is needed for realizing the square lattice simple Hubbard model 
for studying the stripe phases and potential superconducting phases~\cite{Xu_Science2024}. The simple harmonic potential used in this work can be a reasonable approximation for moiré materials whose charge carriers live in one layer, experiencing a modulated potential with long periodicity.
Our results highlight the potential of 2D moiré systems as a quantum simulator, which can complement alternative analog quantum computing platforms such as neutral cold atoms~\cite{Xu_Nature2025}. Compared to the latter, the 2D materials systems offer opportunities for different measurements. 

We employed two theoretical approaches in this study. The first is a direct treatment of the  moir\'e continuum Hamiltonian with DFT, using an LSDA functional tailor-made from
QMC to account for the presence of metallic gates~\cite{Yang_PRB2025}. The second is ab initio downfolding of the 
continuum Hamiltonian into a lattice model. With the first, we compare the 
ground-state phases and properties to what is known in the Hubbard model 
from both UHF~\cite{Xu_JPCM2011,Scholle_PRB2023} and more accurate QMC calculations~\cite{Xu_PRR2022,Chang-PRL-2010} to identify 
potential matching between the continuum system and the model. 
The second approach then provides model parameters for another validation and cross-check. 
It will be valuable to further evaluate the accuracy of DFT/LSDA by benchmark against direct calculations in the continuum by more advanced many-body methods, such as QMC~\cite{Yang_PRL2024,YangYubo_arxiv2024,Xiao_PRR2025}. However,
the cross-check with both UHF and QMC results in the Hubbard model, and 
the consistency between 
our two approaches provide a high level of confidence and allowed us to make quantitative predictions.

\section{Acknowledgments}
Y.Y thanks Chia-Nan Yeh for the help with the Coqui code for downfolding calculations and thanks the
Flatiron Institute for hospitality and computational resources. The Flatiron Institute is a division of the Simons Foundation.

\bibliographystyle{unsrt}
\bibliography{ref.bib}

\end{document}